\documentclass[12pt]{article}
\usepackage{graphics}
\begin{document}
\begin{center}
\large{\textbf{Two electrons in a two dimensional random potential:\\
exchange, interaction and localization}}

\vspace{2mm}
\normalsize
Jorge Talamantes\footnote{email: jtalamantes@csub.edu} \textit{(a)}
and Michael Pollak \textit{(b)}\\
\small
\textit{(a) Department of Physics, California State University,
Bakersfield, CA 93311 USA}\\
\textit{(b) Department of Physics, University of California,
Riverside, CA 92651 USA}
\end{center}
\normalsize
\begin{abstract}
The problem of two electrons in a two-dimensional random potential is
addressed numerically. Specifically, the role of the Coulomb interaction
between electrons on localization is investigated by writing the
Hamiltonian on a localized basis and diagonalizing it exactly. The result of
that procedure is discussed in terms of level statistics, the expectation
value of the electron-electron separation, and a configuration-space inverse
participation number. We argue that, in the interacting problem, a
localization-delocalization crossover in real space does not correspond exactly
to a Poisson-Wigner crossover in level statistics.
\end{abstract}


\vspace{2mm}

\textbf{\S 1. Introduction.}
The problem of two interacting particles (TIP) in a random potential has received
much attention in the last few years. The focus has been primarily on TIP
in one dimension (1\textit{D}), where most
investigators have dealt with particles interacting via an \textit{on-site}
potential. (For a concise summary of the various approaches used, and results
obtained, we refer 
the reader to the articles in ref. \cite{Reviews1D}.) The reason
for all the attention (and controversy) is the result,
first found by Shepelyansky \cite{Shep94}, that the TIP actually 
propagate coherently through a length much larger than the one-particle 
localization radius, which can lead to an enhancement of transport \cite{Imry95}.
We address the related problem of localization in 2\textit{D} systems,
and how it is affected by a \textit{long-range} electron-electron
interaction (EEI).

The more general problem of the interplay between disorder, interaction and quantum
tunneling, and
their combined effect on electronic localization is not new.
While the Hartree Coulomb repulsion introduces an additional random energy
and thus enhances localization, the possibility that quantum correlation due to
the EEI may act to delocalize the electrons was proposed twenty years ago by
Pollak and Knotek \cite{Pollak&Knotek79}, and Pollak \cite{Pollak80},
but a firm answer has not been achieved yet.
Computationally, the main difficulty in the finite-electron-density problem
is the huge phase space for systems of reasonable
size \cite{TPE}. Existing work \cite{TPE,BK,TM,ESV,VES} resorted to various 
approximations. In contrast, the TIP problem for reasonably large systems
can be solved without such approximations: double occupation of sites can be 
accounted for, spin and exchange included, and the entire phase space can be examined.
A motivation for the problem considered here
is the ability to make inferences about approximations made in investigations of
finite-density and few-electron systems
\cite{BS,Emilio,Shep2,Shep:LOC99,Cuevas&Ortuno:LOC99,Shep:PRB2000}.
We hope furthermore that the work may contribute to insight 
into the mechanisms at play in the experimental reports on an observed
metal-insulator transition (MIT) in 2\textit{D} \cite{2dMIT:experiments}.

In the present paper, we deal with two electrons in a random potential interacting
via a long-range Coulomb interaction, and investigate numerically the effect
of that EEI on electronic localization. A tool used frequently to assess 
localization has been the distribution $p(s)$ of nearest-level spacings $s$. In the
absence of interactions, it has been shown \cite{Shklovskii&al} that $p(s)$
shifts from Poisson 
to Wigner as the system goes from being strongly localized to delocalized.
(The salient difference between the two distributions is that Poisson is
maximal as $s \rightarrow 0$, while Wigner vanishes there.)
For \textit{interacting}
systems, such correspondence has never been proven. Still, level spacing
statistics has been used often to study localization also with interactions.
This \textit{may} be reasonable because interaction is not relevant to the logical
connections between the localization $\leftrightarrow$ delocalization 
and Poisson $\leftrightarrow$ Wigner transitions: off-diagonal energies
cause hybridization of site functions and thus delocalization, while at the
same time they increase level repulsion and thus eliminate small level spacings.
This is not to say that the same criteria for the Anderson transition that
were established for non-interacting systems can be \textit{automatically} 
taken over for interacting systems. An interesting case in point is a study of 
the Two-Body Random Interaction Model (TBRIM). Georgeot and Shepelyansky 
\cite{George&Shep97} found that in that model a huge number of non-interacting
eigenstates contribute to the interacting eigenstates
(indicating possible delocalization in real space) even in situations when $p(s)$
is close to Poisson. On the other hand, Jacquod and Shepelyansky \cite{Jacq&Shep97}
used, for the same model, the Wigner-Poisson transition to establish a transition from 
integrability to chaos.

A quantity used commonly to study delocalization in non-interacting 
systems is the inverse participation number $P^{-1}$ \cite{Thouless74}, which
measures over how many sites the (one-particle) wavefunction
spreads. Here, where we allow for interactions, we use an analogous quantity,
the configuration-space inverse participation number $R$ (see ref. \cite{T&P:Jerusalem} 
and eq. (\ref{csipr}) below) which measures over how many configurations the 
many-particle wavefunction spreads. In the limits of strong localization and
of complete delocalization it is easy to see the connection between $P^{-1}$
and $R$: the strong-localization limit, $R=1$, implies the presence of a single
configuration, which in turn implies that each particle is localized on a 
single site. In the complete delocalization limit, $R=1/\tilde{N}$ (where
$\tilde{N}$ is
the total number of configurations), the wavefunction extends uniformly over all
configurations, implying that the particles are uniformly spread over the 
system in real space.

Previous studies on 2\textit{D} random systems with interactions
most relevant to this work
include finite-size scaling of three and four \cite{BS},
and two \cite{Emilio,Shep2,Shep:LOC99,Cuevas&Ortuno:LOC99,Shep:PRB2000} spinless electrons.
All these works have concluded that the
interaction enhances delocalization. In \cite{BS} a \textit{crossover}
from Poisson to Wigner was found, while \cite{Emilio,Shep2,Cuevas&Ortuno:LOC99}
reported a sharp transition with an identifiable critical point. We shall see that 
the enhanced accuracy of this study does not alter the conclusion
that interaction enhances delocalization through most of the energy domain.

In the next two sections, we explain the details of our approach, and describe the
difference with other authors' methods. In the fourth section, we present our results,
and then we offer in the conclusion some final remarks.

\vspace{2 mm}
\textbf{\S 2. The Model.}
In the standard approach (see, \textit{e.g.} refs. \cite{BS,Cuevas&Ortuno:LOC99})
to the problem at hand (sometimes referred to as the Quantum Coulomb Glass 
\cite{ESV}), one considers spinless electrons on sites of a lattice and
represents the Hamiltonian in the basis of local wavefunctions:
\begin{equation}
H_{QCG}=t \sum_{\{ij\}} (a_{i}^{\dagger} a_{j} + h.c.)
  + \sum_{i} \epsilon_{i} a_{i}^{\dagger} a_{i}
  + U \sum_{i>j} (a_{i}^{\dagger} a_{i} a_{j}^{\dagger} a_{j}+
                  a_{i}^{\dagger} a_{j} a_{j}^{\dagger} a_{i})/
    |\vec{r}_{i}-\vec{r}_{j}| \label{H_standard}
\end{equation}
where $\{ij\}$ denotes nearest-neighbor (\textit{n-n}) sites, the operator 
$a_{i}^{\dagger}$ ($a_{i}$) creates (destroys) a (spinless) electron at site
$i$; $\epsilon_i$ is a random site energy chosen in the range $-W/2 < \epsilon_i < W/2$
with uniform probability; $U$ is the strength
of the Coulomb interaction; $\vec{r}$ is a position vector. Both $U$ and $W$ are taken
as independent parameters.

Our model differs primarily in that we include spin, and treat more extensively the
overlaps. The latter is done in two ways: we do not neglect fluctuations
in the \textit{n-n} overlap integrals (due to the differing charge environments
at different pairs of sites), and include overlaps over other than just 
nearest-neighbors (as we cross the MIT from the insulating side, the increase
in the \textit{number} of important distant sites may be more decisive than the
decrease in the \textit{n-n} overlap.)

To accomplish the above we write
the two-electron wavefunctions in terms of a basis set of appropriately
symmetrized products of the one-particle local wavefunctions
\begin{eqnarray}
\Phi_{ij} & = & \phi_{ij}^{(S)} \times \sigma^{(A)} \mathrm{,  and} \label{singlets} \\
\Phi_{ij} & = & \phi_{ij}^{(A)} \times \sigma^{(S)} \mathrm{,} \label{triplets}
\end{eqnarray}
where $i,j$ are site labels, $\phi_{ij}$ ($\sigma$) is the spatial (spin) part of 
the two-electron configuration, and 
the superscripts $S$ and $A$ indicate the symmetry of the wavefunction under particle
exchange. The
$\phi_{ij}$ are constructed in the usual manner, \textit{i.e.} by symmetrizing
(or antisymmetrizing) products
such as $\varphi_{i}(1) \varphi_{j}(2)$ of one-electron orbitals (centered on $i$ and $j$)
$\varphi$ for electrons 1 and 2.
Double occupation of sites comes in through $\phi_{ii}^{(S)}$. Clearly, the
$\phi_{ij}^{(S)}$ ($\phi_{ij}^{(A)}$) are singlet (triplet) configurations.
We take $\varphi \sim \exp(-r/a_B)$,
with $r$ the electron-core distance and $a_B$ the microscopic (Bohr) radius corresponding
to those orbitals.

We write the Hamiltonian
\begin{eqnarray}
H&=&\sum_{\alpha=1}^{2} (T_\alpha + V_\alpha+\varepsilon_\alpha)+
  \frac {e^{2}}{\kappa r_{12}}, \mathrm{with} \label{hamiltonian}\\
T_\alpha&=&-\frac{\hbar^{2}}{2m^{*}} \nabla^{2}_\alpha, \nonumber \\
V_\alpha&=&-e \cdot q_s \sum_{i=1}^{n_s} \frac{1}{\kappa r_{i \alpha}} \mathrm{,  and} \nonumber \\
\varepsilon_\alpha&=&\sum_{i=1}^{n_s} \epsilon_i \delta(\vec{r}_\alpha-\vec{r}_i), \nonumber
\end{eqnarray}
where $\alpha$ labels the electrons, $T$, $V$, and $\varepsilon$ are the operators 
corresponding to the
kinetic energy, the interaction of the electrons with the cores, and the random potential, respectively;
$e$ is the electronic charge; $\kappa$ is the dielectric constant; $r_{12}$
is the electron-electron (\textit{e-e}) distance; $m^*$ is the electron mass; $q_s$ is the charge
on a site; $n_s$ is the number of sites in the system;
$\epsilon_i$ is a random energy chosen as in eq. (\ref{H_standard}),
with $W$ equal to the
\textit{n-n} Coulomb energy $e_C$ (\textit{i.e.} we set $W=U=1$ in units of a \textit{n-n}
Coulomb energy). We vary the site concentration by changing the
\textit{n-n} distance $r_{nn}$ as a parameter.

Analytic solutions for 
$\langle \phi_{ij}^{(S)} | H | \phi_{kl}^{(S)} \rangle$ and
$\langle \phi_{ij}^{(A)} | H | \phi_{kl}^{(A)} \rangle$
were derived and written in terms of one-, two-, three-, and four-center integrals.
The procedure is very tedious but straightforward. The equations are not given here
for reason of space, and because the TIP case is not general;
the equivalent expressions for systems with an arbitrary number of spinless fermions
were published in \cite{T&P:Jerusalem}.
The integrals were performed numerically for processes $\{ij\} \rightarrow \{kl\}$ which
involved \textit{(i)} no electron transfers (\textit{i.e.} diagonal matrix elements);
or \textit{(ii)} a one-electron \textit{n-n} transfer;
or \textit{(iii)} a next \textit{n-n} transfer;
or \textit{(iv)} a next-to-next \textit{n-n} transfer;
or \textit{(v)} two simultaneous \textit{n-n} transfers.
All other off-diagonal matrix elements were set to zero.
$H$ does not
depend on spin, so $\sigma$ enters into the picture only by dictating the
symmetry (under particle exchange) of the corresponding $\phi_{ij}$.
Naturally, 
the off-diagonal matrix elements between $\phi_{ij}$ of different symmetry 
are set to zero; thus the matrix $H$ splits into two blocks which we diagonalize
separately using standard techniques. The result is the set of eigenenergies
$E_I$, and the corresponding states $\Phi_I$ in which the 
$\phi_{ij}$ come in with amplitudes $A_{I,ij}$:
\begin{equation}
\Phi_I^{(S)}=\sum_{\{ij\}} A_{I,ij}^{(S)} \phi_{ij}^{(S)}, \textrm{ and }
\Phi_I^{(A)}=\sum_{\{ij\}} A_{I,ij}^{(A)} \phi_{ij}^{(A)}. \label{eigenstates}
\end{equation}
For simplicity we refer below to $\Phi_I^{(S)}$ as ``singlets" and 
$\Phi_I^{(A)}$ as ``triplets".

In this work we take $W$ to
always equal $e_C$; thus, $W=e_C \sim 1/r_{nn}$. Since $r_{nn}$ is a parameter here, $W$
changes accordingly. We do this because then the effect of interaction
is most important, and also because this condition prevails in many experiments,
for example, in the vast experimental literature on impurity conduction at moderate compensation
\cite{Mott67}. In impurity conduction $W$ corresponds to the Coulomb energy over the 
mean \textit{n-n} majority-minority ion distance, while $U$ corresponds to the
Coulomb energy over the mean carrier-carrier separation. The two are quite similar 
unless the compensation is either very small or very large.

The present model also differs from other models in how it deals with electrical
neutrality. In most works neutrality is achieved by placing a compensating
charge $q_s=+|e| n_e/n_s$ on every site.
However, since the microscopic radius of the electrons is usually
determined by the charge $q_s$ of the core, this radius would strongly depend on
how electrical neutrality is established. (For the TIP problem $n_e$ is fixed,
but $n_s$ is generally taken as a parameter; thus,
different models for compensation may in fact yield different results.) To avoid
such problems, this work does not consider the compensating charge explicitly. To
make contact with experimental situations, we place a core charge of magnitude $+|e|$
at each site. The macroscopic radius is then determined by an effective mass
and a dielectric constant. Since there are more sites than carriers,
the system is not neutral.
A question we need to answer is whether the properties of interest here, 
\textit{i.e.} the degree of localization and its dependence on various parameters,
are \textit{fundamentally} affected by the lack of neutrality. To answer this question
we consider whether one can construct \textit{some} model for the compensating
charge such that it has no effect on the results of the computations as far as the
localization properties are concerned. The answer is that such a model exists, namely
a spatially uniform distribution of the compensating charge (which may be placed on
a separate plane, not unlike the situation in gated devices). This charge merely
contributes two constant energies: a self-interaction, and the interaction with the
two electrons. We may thus conclude
that the lack of neutrality \textit{per se} does not affect our results. However,
it should be borne in mind that in reality the particular way in which charge neutrality
is realized may affect somewhat the localization properties.

\vspace{2 mm}
\textbf{\S 3. Procedure.} We set up ``samples" on the computer with $n_s$
($=L \times L$) sites arranged on a lattice,
and diagonalize eq. (\ref{hamiltonian}) for the parameters $L$ and $r_{nn}$.
For definiteness, we take $a_B$ and $\kappa$ to be 10\AA\ and 3
respectively, which yields and effective mass $m^*=0.16m$, with
$m$ the electron mass -- these values seem appropriate for 2\textit{D} systems.
Cyclic boundary conditions are used. In what follows, we use $a_B$ as the unit
of distance. In every case the number of samples was sufficient to obtain
at least $1.5 \times 10^4$ levels for each pair $\{n_s,r_{nn}\}$.

As one measure of localization, we compute $p(s)$.
(We dropped up to 100 states from the band edges.) 
As in \cite{Emilio}, a parameter
$\eta=(\mathrm{var}(s)-0.273)/(1.0-0.273)$ is computed as a measure of how close
$p(s)$ is to a Poisson ($\eta=1$) or Wigner ($\eta=0$) distribution.
As an alternate measure of localization we compute the
configuration-space inverse participation number as in ref.~\cite{T&P:Jerusalem}:
\begin{equation}
R_I=\sum_{\{ij\}} |A_{I,ij}|^4. \label{csipr}
\end{equation}
In addition to $p(s)$ and $R_I$ we examine the behavior of the quantity
\begin{equation}
\lambda_{I} = \sum_{\{ij\}} r_{ij} |A_{I,ij}|^2, \label{corr_length}
\end{equation}
which is the expectation value of the \textit{e-e}
separation when the system is
in state $\Phi_I$.
This expectation value is computed here
for a direct glimpse at the behavior of the $\Phi_I$ in \textit{real space}.
One expects that
in the localized regime $\lambda_I$ is strongly correlated with $E_I$ 
due to the EEI (larger
$\lambda_I$ correspond to smaller $E_I$), whereas $\lambda_I$ should become
essentially independent of $E_I$ as configuration-mixing increases.
In the absence of interactions, such correlations should of course not
be present.

To measure the importance of secondary tunneling processes, we
investigate the following quantities: \textit{(i)} the width $w$, and the average
$t_{nn}$ of the distribution of off-diagonal elements corresponding to \textit{n-n}
processes; \textit{(ii)} the sum $S^{(1)}$ of all off-diagonal elements corresponding
to \textit{n-n} processes; \textit{(iii)} a similar sum $S^{(\sqrt{2})}$ for all next 
\textit{n-n} processes; and \textit{(iv)} a sum $S^{(2)}$ of such matrix elements 
which correspond to either next-to-next \textit{n-n} or two simultaneous one-electron \textit{n-n}
transfers. We note that the sums $S^{(\cdots)}$ account in a crude way for not
only the magnitude of the off-diagonal matrix elements, but also for the
size of the phase space occupied by those processes (\textit{e.g.} even though
a matrix element corresponding to a \textit{n-n} transition may be 
significantly larger than one corresponding to two simultaneous \textit{n-n}
transfers, the number of matrix elements of the latter
type is much larger -- and thus such processes may contribute
significantly to coherent tunneling). Also, we note that the
standard approach assumes $t_{nn} \gg w$, $S^{(1)} \gg S^{(\sqrt{2})}$ and 
$S^{(1)} \gg S^{(2)}$.

\vspace{2 mm}
\textbf{\S 4. Results and discussion.}
Double occupation of sites does not, of course, occur in the triplets case,
and a single band of eigenenergies results from the diagonalization
procedure. The width of this band naturally increases with decreasing 
$r_{nn}$ because the configurations hybridize, and level repulsion
increases. For singlets, however, at large $r_{nn}$ two such bands separated
by a gap result: a lower band arises from hybridization of configurations in which sites 
are singly occupied (to which we will refer as type-\textit{s} configurations)
-- as in the triplets case,
and an upper band arises from configurations in which sites are doubly-occupied
(type-\textit{d} configurations), which for the most part do not hybridize;
with decreasing $r_{nn}$, the gap narrows as (mostly type-\textit{s}) configurations 
hybridize, and level repulsion increases. As $r_{nn}$ is decreased further,
type-\textit{d} configurations start to mix with each other, and with
type-\textit{s} configurations, until eventually (for $r_{nn} \leq \sim 4$),
the gap actually disappears, and the two bands start to join together.
We present results here for $r_{nn}=4 \textrm{-} 12$. In what follows, except
for our discussion
of $\lambda_I$, our results refer to the triplets band, and to the singlets'
lower band.

\begin{figure}
\includegraphics[10mm,10mm][117mm,75mm]{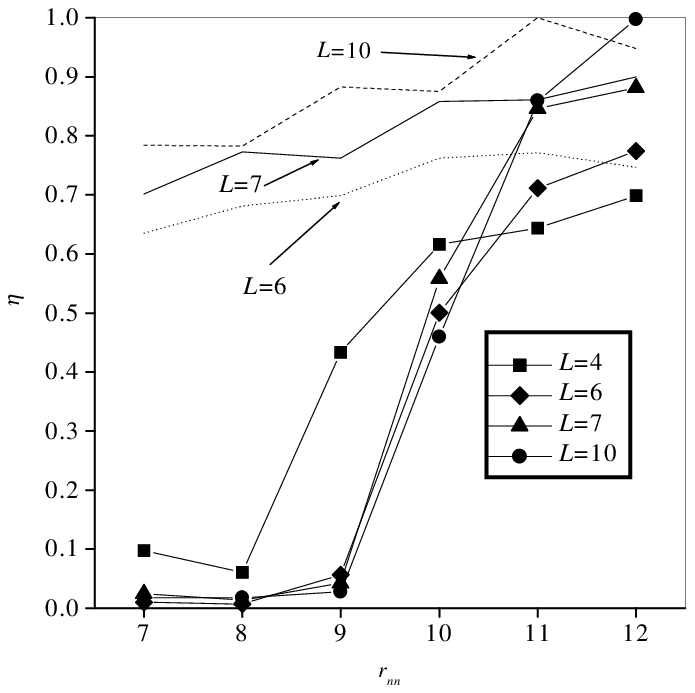}
\caption{Distribution $p(s)$ of nearest level spacing. The lines
without symbols correspond to the non-interacting Hamiltonian.
The lines with symbols correspond
to the full Hamiltonian [eq. \ref{hamiltonian}].}
\label{eta}
\end{figure}

We first present the basic conclusions regarding the effect of the EEI on localization.
Fig. \ref{eta} shows a comparison of $\eta (r_{nn})$ between the interacting and 
non-interacting systems [\textit{i.e.} eq. (\ref{hamiltonian}) without the last
term]. It is very clear that this criterion shows EEI to enhance
delocalization -- while with EEI delocalization occurs at $r_{nn} \sim 10$,
without EEI delocalization is out of the range of the figure, namely somewhere
at $r_{nn}<7$. We note that there is no clear small-size scaling behavior, suggesting
a crossover ($r_{nn} \sim 9 \textrm{-} 11$),
rather than a critical transition. This is in agreement with 
\cite{BS}, but differs from \cite{Emilio,Shep2,Cuevas&Ortuno:LOC99,Shep:PRB2000}.
The differences may be due to differences in models
and choice of parameters \cite{screening}. 

\begin{figure}
\includegraphics[10mm,10mm][117mm,195mm]{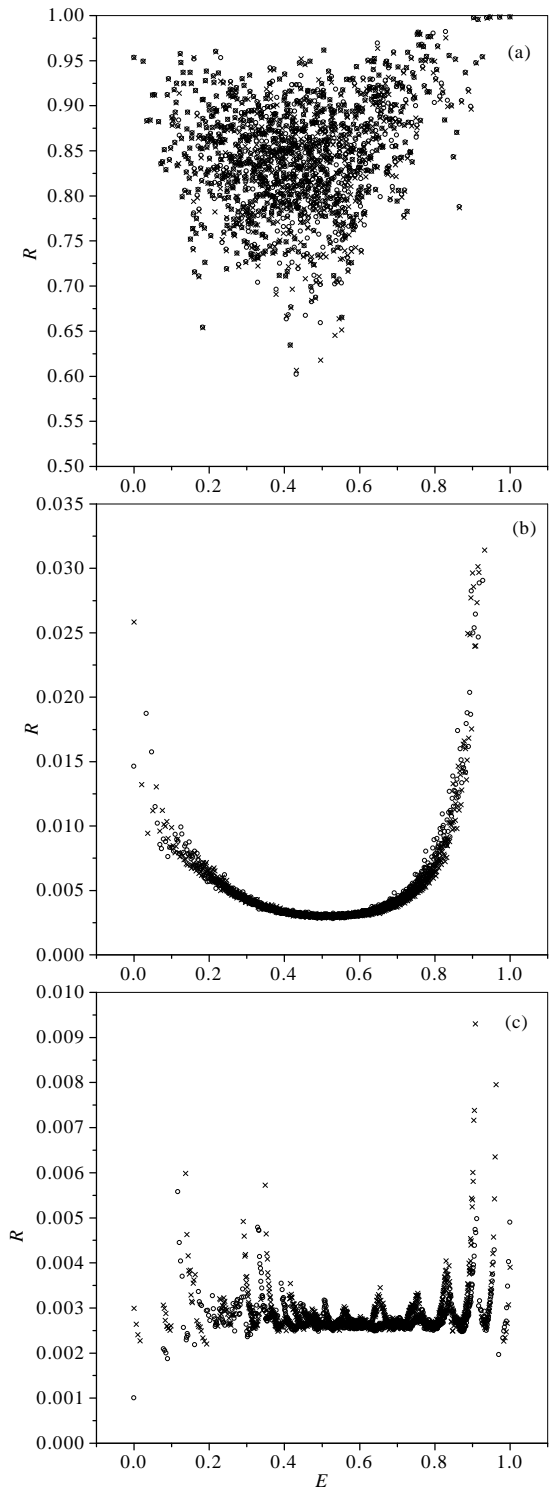}
\caption{ Inverse participation number in configuration space $R$ with $L=7$
for singlets ($\circ$) and triplets ($\times$),
and (a) $r_{nn}=12$; (b) $r_{nn}=8$; (c) $r_{nn}=6$. The energy scale is
described in the text.}
\label{R:sandt}
\end{figure}

\begin{figure}
\includegraphics[10mm,10mm][117mm,75mm]{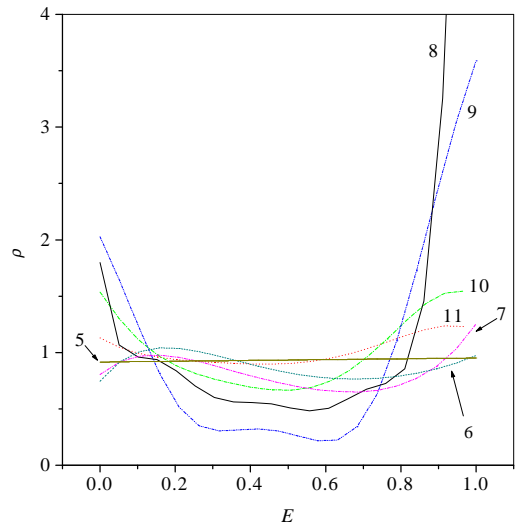}
\caption{ The ratio $\rho$ of the configuration-space inverse participation number $R$
with EEI to $R$ without. The numbers on the graph indicate the
corresponding values of $r_{nn}$. $\rho=1$ for $r_{nn}=12$ (not shown).
The energy scale is described in the text.}
\label{R:EEI}
\end{figure}

We present in figs. \ref{R:sandt} and \ref{R:EEI} our results for $R_I$ with
$L=7$, but we checked that the plots for 
other values of $L$ are qualitatively the same.
To evaluate the importance of spin, we plot in fig. \ref{R:sandt} the number $R_I$ for
singlets and triplets.
In both cases, the energies are measured from the bottom of the band, and
have been normalized by the energy difference between the top and the bottom
of the band.
The most noticeable features in fig. \ref{R:sandt} are:
1) in the localized regime [fig. \ref{R:sandt}(a)] there is no discernible
difference between singlets and triplets;
2) For $r_{nn}$ close to the transition and into the delocalized regime,
the lowest energy states of the singlets are \textit{always} more delocalized
than the lowest energy triplets. This is exemplified by figs. \ref{R:sandt}(b)
and \ref{R:sandt}(c). This difference between the singlets and the triplets 
seems to be larger the smaller the $r_{nn}$. 3) In the well-delocalized regime
there appear intriguing fluctuations of $R_I$ with energy, for both singlets
and triplets, as exemplified by fig. \ref{R:sandt}(c).
The fluctuations are not random -- there is a definite correlation in $R_I$
over a finite distance in energy. This persists from one realization of
random energies to another, and persists also for increased $L$,
though with a different characteristic correlation width. Possibly the
fluctuations are connected with interference effects that come into play,
due to our use of cyclic boundary conditions, once the extent of the wavefunction
spans the entire system. Interestingly, similar fluctuations were observed
\cite{Barelli&al96} in a real-space inverse participation ratio for
TIP in the Harper model where the non-interacting eigenstates are delocalized.

In fig \ref{R:EEI}, $\rho$ is the
average (over singlets and triplets) $R_I$ with EEI to the average $R_I$ without
the interactions, \textit{i.e.} $R_I$ is computed with and without 
the last term in eq. (\ref{hamiltonian}),
and the results are averaged over singlets and triplets
-- $\rho$ is given by the ratio of the two averages. Thus, $\rho < 1$ indicates
delocalization (in configuration space) induced by the EEI, whereas
$\rho > 1$ points to interaction-induced localization.
Over most of the regime of $E_I$ the
plots confirm the previous result about the effect of EEI on delocalization.
However, at the extremes of the energy regime, the opposite
appears to happen -- the wavefunctions for the interacting system are more
localized, at least in configuration space.
(It has been shown \cite{Pollak:Rackeve} that EEI can,
under certain conditions enhance either localization or delocalization, and this
in fact has been found to be the case \cite{VES}.)
We discuss the enhanced localization at the band edges in the context of
\cite{Pollak:Rackeve}. The enhancement or supression of localization depends 
on the general effect of the EEI on the ratios
$H_{IJ}/D_{IJ}$, where $H_{IJ} \equiv \langle \Phi_I | H |\Phi_J \rangle$ and
$D_{IJ} \equiv \langle \Phi_I|H|\Phi_I \rangle - \langle \Phi_J|H|\Phi_J\rangle$. 
[Configuration indices $I$, $J$ here replace the pair indices $i$, $j$ of
eqns. (2) and (3).] If the EEI enhances $H_{IJ}/D_{IJ}$ it supresses localization, in the
opposite case it enhances it. $H_{IJ}$ and $D_{IJ}$ are both separately enhanced by the EEI
\cite{Pollak:Rackeve}, so the effect on localization depends on which is
enhanced more. Now consider, for example, the lower band edge, \textit{i.e.}
the very low-energy end of the spectrum. The level spacing, and thus $D_{IJ}$,
is enhanced by the interaction because it excludes configurations with nearby
electrons. The enhancement of $H_{IJ}$ by interactions comes to a large degree
from correlated \textit{n}-particle (here 2-particle) transitions, but this 
effect is important mainly when the electrons are reasonably close-by. A
somewhat parallel argument can be made for the upper band edge. It is 
interesting to note that similar effects of enhanced localization were reported
in \cite{Barelli&al96} for rather different systems.

\begin{figure}
\includegraphics[10mm,10mm][117mm,140mm]{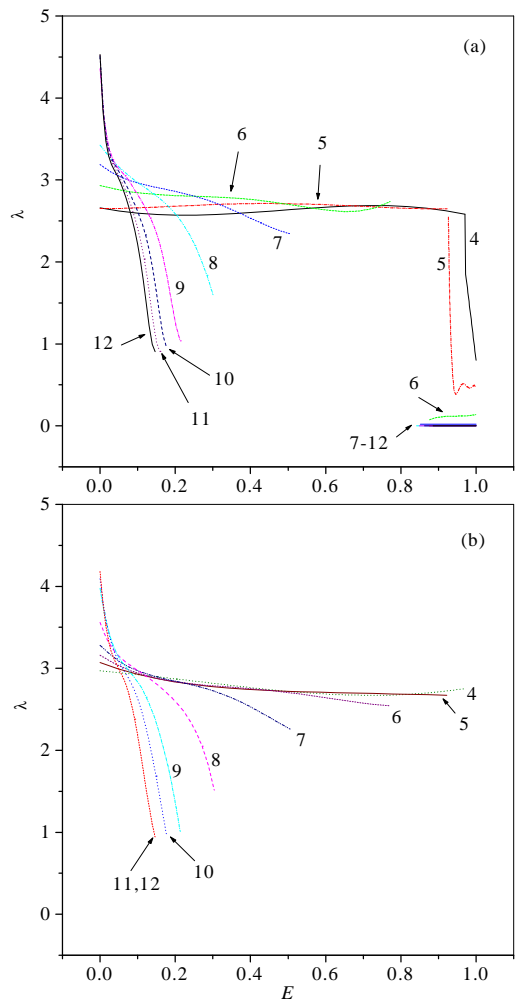}
\caption{Fits for $\lambda (E)$ for (a) singlets, and
(b) triplets with $L=7$. The numbers on the graph correspond to the values of $r_{nn}$.
In (a) the upper bands coincide for
$r_{nn}=7 \textrm{-} 12$; in (b) the plots for $r_{nn}=11,12$  coincide.
The energy scale is described in the text.}
\label{lambda}
\end{figure}

We consider next the question of \textit{e-e} separation due to interaction
as measured by $\lambda_I$. Without the EEI the positions of the two electrons 
are uncorrelated so $\lambda_I$ is expected to vary at random from configuration
to configuration and we ascertained that this is indeed the case. Fig.
\ref{lambda} shows polynomial fits for $\lambda_I (E_I)$ with EEI for $L=7$.
(Again, we checked that the results for other values of $L$ are qualitatively
the same.)
Fig. \ref{lambda}(a) corresponds to singlets, and fig. \ref{lambda}(b) to triplets.
The numbers on the graph refer to the values of $r_{nn}$. In the singlets
case we now include both bands.
This figure is similar to results presented in \cite{TPV:Murcia},
but here we present an average over many realizations of the random potential.
In figs. \ref{lambda}(a) and (b) the energies are measured from the bottom
of the lower band. In fig. \ref{lambda}(a) the energies are normalized to the
energy difference between the top of the upper band the the bottom of the
lower band; for comparisons between figs. \ref{lambda}(a) and (b), the
triplets' energies have been normalized so that the top of their band coincides
with the top of the singlets' lower band.
The results shown in the figure can be conveniently discussed by dividing
the plots into three groups:
1) The singlets [fig. \ref{lambda}(a)] include states at the high energy end of the
figure which do not have a triplet counterpart [fig. \ref{lambda}(b)].
As discussed above, these states
correspond to double occupation of sites: for large $r_{nn}$, $\lambda_I$ is quite
small and the energy large, as expected for two electrons residing on the same site.
As $r_{nn}$ decreases, some states still have the large energy, but now $\lambda_I$
is larger (as is evident from the steep lines with $r_{nn}=4,5$).
These are likely to be states
with a reasonably large component of a doubly-occupied site, hybridized with
configurations where electrons are more remote from each other.
The other features of fig. \ref{lambda} are common to the triplets and the
singlets and correspond to states where double occupation of sites is minimal
or non-existent.
2) The curves representing large values of $r_{nn}$ (small overlaps) are quite steep,
\textit{i.e.} $\lambda_I$ decreases sharply with increasing $E_I$. This is easily
understandable as an increase in the repulsion energy with a decreasing distance
between occupied sites, \textit{i.e.} decreasing $\lambda_I$.
3) The dependence of $\lambda_I$ on $E_I$ weakens as $r_{nn}$ decreases and 
configuration-mixing increases -- $\lambda_I$
becomes nearly independent of $E_I$ and quite large for $r_{nn}=5$. The crossover takes place around
$r_{nn}=6 \textrm{-} 9$ in both fig. \ref{lambda}(a) and (b).

We note that the crossover in $\lambda_I$ is lower than in fig. \ref{eta}.
In this model, $W/t = e^{r_{nn}}/[r_{nn} (1+r_{nn})]$, so a small difference
in $r_{nn}$ implies a comparatively larger one in $W/t$. This gives $W/t \sim 3,5,10,20,40,
90,200,500,1000$ for $r_{nn}=4,5,6,7,8,9,10,11,12$, respectively. One must be careful, 
however, when comparing our results with those of other authors because here
$W/t=e_C/t$.

\begin{figure}
\includegraphics[10mm,10mm][117mm,75mm]{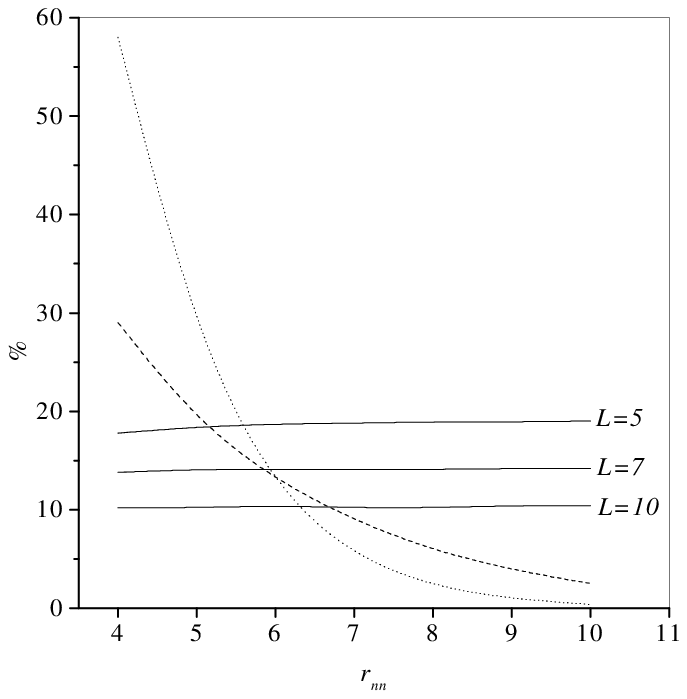}
\caption{An evaluation of the importance of secondary processes
relative to \textit{n-n} tunneling. The solid lines are plots of $w/t_{nn}$.
The dashed line is $S^{(\sqrt{2})}/S^{(1)}$ \textit{vs.} $r_{nn}$
(the same plot for $L=5,7,10$), and the dotted line is $S^{(2)}/S^{(1)}$
\textit{vs.} $r_{nn}$ (also the same plot for $L=5,7,10$). The vertical axis
is given as a percent in all cases.}
\label{odestats}
\end{figure}

In fig. \ref{odestats} we present our analysis of the importance of fluctuations
in \textit{n-n} hopping energies, and of secondary 
hopping processes. First we note that the relative width of the distribution
of \textit{n-n} hopping energies remains roughly constant with $r_{nn}$, and
decreases with system size. As the size of the phase space increases with $n_s$,
the relative importance of these fluctuations decreases, and probably disappears
in the thermodynamic limit. Furthermore, secondary hopping processes are not
very important in the localized regime; however, they are important
(as expected) in the delocalized regime. These processes become significant
for $r_{nn} \leq \sim 6$, and in fact, next-to-next \textit{n-n}
and simultaneous \textit{n-n} processes are more important than next \textit{n-n}
in the delocalized regime. We surmise that these secondary processes
``push" the transition towards higher values of $r_{nn}$.

We now turn to the specific aspects learned from those features of our model
which go beyond other works \cite{Emilio, Shep2, Shep:LOC99, Cuevas&Ortuno:LOC99},
namely the inclusion of spin and
the more general consideration of elastic tunneling. We note that:
1) As expected, spin plays no
role deep in the localized regime (since the exchange energy is proportional 
to $\langle \varphi_i | \varphi_j \rangle^2$). More unexpectedly, 
spin also seems to have little importance near the crossover to
delocalization. Perhaps spin plays a larger role when one considers more than
two electrons. 
2) Type-\textit{d} configurations
play little or no role near the crossover -- they
become important only for $r_{nn} \leq \sim 4$.
3) In the localized extreme (\textit{i.e.} $r_{nn} \geq 12$), the rather short span
in energy of the eigenstates comes from the random and Coulomb energies;
as $r_{nn}$ decreases,
the broadening of the range of eigenenergies is
attributable to the growing off-diagonal energies of $H$.
4) The standard model underestimates the delocalizing effect of the EEI
for small $r_{nn}$ especially, as shown in fig. \ref{odestats}. Many-electron
coherent processes are probably important in the finite-electron-density
problem. This phenomenon was first pointed out in \cite{TM}. Secondary
processes are important especially in the delocalized regime -- one should
consider them if investigating systems around the MIT.

\textbf{\S 5. Conclusions.}
The model used here yields zero density in the thermodynamic limit, and so no definite
claims or comparisons with experiments can be made. However, if we take the electronic 
localization-delocalization transition at $r_{nn}=6$, this simple model yields 
the critical concentration $n_c=10^{11} \textrm{ cm}^{-2}$. Whereas our systems are
quite different from the experimental ones (see, \textit{e.g.} \cite{K&DasS},
where a 2\textit{D} MIT was reported for high-mobility systems at low temperatures),
it is interesting to note that the experimental value of $n_c$ is the same as here.
This of course might be just fortuitous.

Where collective hopping of the two electrons is coherent 
($r_s \leq \sim 7$),
$\lambda$ can be interpreted as a coherence length.
It is of interest to note that a
crossover of $p(s)$ from Poisson to Wigner occurs at a somewhat larger $r_{nn}$
than the crossover from a large variation in $\lambda_I (E_I)$ to $\lambda_I (E_I) \sim
\mathrm{const.}$ We have included in our level spacing analysis most of the eigenstates of 
eq. (\ref{hamiltonian}),
as opposed to only states in the middle of the band, which is the customary procedure
(see \textit{e.g.} \cite{BS,Emilio,Cuevas&Ortuno:LOC99}). Since we are including
states which tend to be more localized than those in the middle of the band
(as is evident from fig. \ref{R:EEI}),
the effect is that our level spacing statistics picture is skewed towards the
\textit{localized} regime.
The disagreement between the crossover on $p(s)$ and that in $\lambda_I$ would
be stronger if we were to follow standard practice.
We believe that (unlike in the non-interacting case) for interacting systems, delocalization
in \textit{real space} requires a somewhat larger overlap for \textit{n-n} sites than does the
transition to Wigner statistics. This is in keeping with previous work
\cite{T&P:Jerusalem,T&P:Vancouver}, where it was observed
that the wavefunctions are ``swiss cheese-like" (\textit{i.e.} not compact
in real space) without EEI, but space-filling with the
interactions \cite{cheese}; therefore, while the EEI may make the wavefunction
extend over more sites, it
does not similarly increase its \textit{spatial} extent -- $\lambda_I$ might
require a larger overlap than $p(s)$ (for a
crossover to take place) because the EEI makes the wavefunction more
compact in real space. In a sense, there is a similarity here with \cite{George&Shep97},
where it was shown that interacting eigenstates may contain contributions from many
non-interacting eigenstates, even when $p(s)$ is close to Poisson. The difference is that we get 
$p(s)$ close to Wigner in some situations when one might expect Poisson. For example,
from fig. \ref{lambda}, at $r_{nn}=9$ one would expect $p(s)$ similar to Poisson since $\lambda_I$
depends strongly on $E_I$; however, fig. \ref{eta} reveals that $p(s)$ is nearly Wigner.
We attribute the discrepancy between this work and the result in \cite{George&Shep97}
to the different models used.

It is clear that $p(s)$ is Poisson well inside the localized regime, and Wigner
well into the delocalized regime; however, the crossover in level statistics
and its relationship to a localization-delocalization transition in \textit{real space}
is not as well established as it is in the non-interacting problem, and requires
further study. Furthermore, 
whereas it is clear that deep localization in configuration space
implies deep localization in real space, and the same holds true for well-delocalized systems,
the actual transition need not occur simultaneously in the two domains.

\vspace{2mm}
\textbf{Acknowledgements.} The support of the National Science Foundation under grant
DMR-9803686 is gratefully acknowledged, as are conversations with I. Varga, and
some useful comments by the (unfortunately anonymous) second referee of this paper.

%

\pagebreak

\begin{thebibliography}{99}
\bibitem{Reviews1D}
K. M. Frahm, Eur. Phys. J. B \textbf{10}, 371 (1999);
R. R\"omer, M. Leadbeater and M. Schreiber, Ann. Phys. \textbf{8}, 675 (1999).
\bibitem{Shep94}
D. Shepelyansky, Phys. Rev. Lett. \textbf{73}, 2607 (1994).
\bibitem{Imry95}
Y. Imry, Europhys. Lett. \textbf{30}, 405 (1995).
\bibitem{Pollak&Knotek79}
M. Pollak and M. L. Knotek, Journal of Non-Cryst. Sol. \textbf{32}, 141 (1979).
\bibitem{Pollak80}
M. Pollak, Phil. Mag. B \textbf{42}, 781 (1980).
\bibitem{TPE}
J. Talamantes, M. Pollak and L. Elam, Europhys. Lett. \textbf{35}, 511 (1996).
\bibitem{BK}
D. Belitz and T. R. Kirkpatrick, Rev. Mod. Phys. \textbf{66}, 261 (1994).
\bibitem{TM}
J. Talamantes and A. M\"{o}bius, phys. stat. sol. (b) \textbf{205}, 45 (1998).
\bibitem{ESV}
F. Epperlein, M. Schreiber and T. Vojta, Phys. Rev. B \textbf{56}, 5890 (1997).
\bibitem{VES}
T. Vojta, F. Epperlein and M. Schreiber, Phys. Rev. Lett. \textbf{81}, 4212 (1998).
\bibitem{BS}
R. Berkovits and B. I. Shklovskii, J. Phys.: Condens. Matter \textbf{11}, 779 (1999).
\bibitem{Emilio}
E. Cuevas, Phys. Rev Lett. \textbf{83}, 140 (1999).
\bibitem{Shep2}
D. Shepelyansky, cond-mat/9905231 \textit{preprint} (1999).
\bibitem{Shep:LOC99}
D. Shepelyansky, Ann. Phys \textbf{8}, 665 (1999).
\bibitem{Cuevas&Ortuno:LOC99}
E. Cuevas, and M. Ortu\~no, Ann. Phy. \textbf{8}, Spec. Issue SI 37 (1999).
\bibitem{Shep:PRB2000}
D. Shepelyansky, Phys. Rev. B \textbf{61}, 4588 (2000).
\bibitem{2dMIT:experiments}
S. V. Kravchenko, G. V. Kravchenko, J. E. Furneaux, V. M. Pudalov and D. M. D'Iorio, Phys. Rev. B \textbf{50}, 8039 (1994);
S. V. Kravchenko, W. E. Mason, G. E. Bowker, J. E. Furneaux, V. M. Pudalov and M. D'Iorio, Phys. Rev. B \textbf{51}, 7038 (1995);
S. V. Kravchenko, D. Simonian, M. P. Sarachik, W. E. Mason and J. E. Furneaux, Phys. Rev. Lett. \textbf{77}, 4938 (1996);
D. Simonian, S. V. Kravchenko and M. P. Sarachik, Phys. Rev. B \textbf{55}, R13 421 (1997);
D. Simonian, S. V. Kravchenko, M. P. Sarachik and M. V. Pudalov, Phys. Rev. Lett. \textbf{79}, 2304 (1997);
D. Popovi\'{c}, A. B. Fowler and S. Washburn, Phys. Rev. Lett. \textbf{79}, 1543 (1997).
\bibitem{Shklovskii&al}
see, \textit{e.g.} B. Shklovskii, B. Shapiro, B. R. Sears, P. Lambrianides and
H. B. Shore, Phys. Rev. B \textbf{47}, 11487 (1993).
\bibitem{George&Shep97}
B. Georgeot and D. L. Shepelyansky, Phys. Rev. Lett. \textbf{79},4365 (1997).
\bibitem{Jacq&Shep97}
Ph. Jacquod and D. L. Shepelyansky, Phys. Rev. Lett. \textbf{79},1837 (1997).
\bibitem{Thouless74}
D. J. Thouless, Physics Reports \textbf{13}, 93 (1974).
\bibitem{T&P:Jerusalem}
J. Talamantes and M. Pollak in \textit{Hopping and Related Phenomena}, edited by O. Millo and 
Z. Ovadyahu (Racah Institute of Physics, Jerusalem) 1995.
\bibitem{Mott67}
see, \textit{e.g.}, N. F. Mott, Adv. Phys. \textbf{16}, 49 (1967).
\bibitem{screening}
It should be pointed out that any conclusions reached from finite-size scaling
behavior of computer size samples cannot be applied to experimental size samples
because of self-screening. Without screening there cannot be a thermodynamic limit -- but
screening \textit{per force} introduces into the problem a screening length, which
precludes self-similarity and thus eliminates the possibility of a simple critical
transition.
\bibitem{Barelli&al96}
A. Barelli, J. Bellisard, Ph. Jacquod, D. Shepelyansky, Phys. Rev. Lett.
\textbf{77}, 4752 (1996).
\bibitem{Pollak:Rackeve}
M. Pollak, phys. status sol. (b) \textbf{205}, 35 (1998).
\bibitem{TPV:Murcia}
J. Talamantes, M. Pollak, and I. Varga, phys. stat. sol. (b) \textbf{218}, 119 (2000).
\bibitem{K&DasS}
T. M. Klapwijk and S. Das Sarma, Solid State Communications \textbf{110}, 581 (1999).
\bibitem{T&P:Vancouver}
J. Talamantes and M. Pollak, \textit{Physics of Semiconductors},
edited by D. J. Lockwood (World Scientific Publishing, Singapore) 1995.
\bibitem{cheese}
The idea is that the EEI pushes electrons away from each other, thus clearing
out the vicinity of occupied sites. This enhances the coherent transfer of
electrons to nearby sites. This enhancement is absent in the non-interacting case.
\end{thebibliography}
\end{document}